\documentclass{article}

\usepackage{microtype}
\usepackage{graphicx}
\usepackage{subcaption}
\usepackage{booktabs} 
\usepackage{tabularx} 

\usepackage{hyperref}

\usepackage[preprint]{icml2026}

\usepackage{amsmath}
\usepackage{amssymb}
\usepackage{mathtools}
\usepackage{amsthm}

\usepackage{algorithm}
\usepackage{algorithmic}
\usepackage{bm}

\usepackage[capitalize,noabbrev]{cleveref}

\theoremstyle{plain}

\theoremstyle{definition}

\theoremstyle{remark}

\usepackage[textsize=tiny]{todonotes}

\icmltitlerunning{AdaptiveLoad: Towards Efficient Video Diffusion Transformer Training}

\begin{document}

\twocolumn[
  \icmltitle{AdaptiveLoad: Towards Efficient Video Diffusion Transformer Training}



  \icmlsetsymbol{equal}{*}

  \begin{icmlauthorlist}
    \icmlauthor{Yucheng Guo}{equal,JDT AI Infra}
    \icmlauthor{Yongjian Guo}{equal,Tsinghua University,JDT AI Infra}
    \icmlauthor{Zhong Guan}{Tianjin University,JDT AI Infra}
    \icmlauthor{Haoran Sun}{Peking University,JDT AI Infra}
    \icmlauthor{Wen Huang}{Tsinghua University,JDT AI Infra}
    \icmlauthor{Wanting Xu}{JDT AI Infra}
    \icmlauthor{Jing Long}{Peking University,JDT AI Infra}
    \icmlauthor{Shuai Di}{JDT AI Infra}
    \icmlauthor{Junwu Xiong}{JDT AI Infra}
  \end{icmlauthorlist}

  \icmlaffiliation{Tsinghua University}{ Tsinghua University, Beijing, China}
  \icmlaffiliation{Peking University}{Peking University, Beijing, China}
  \icmlaffiliation{Tianjin University}{Tianjin University, Tianjin, China}
  \icmlaffiliation{JDT AI Infra}{JDT AI Infra, Beijing, China}

  \icmlcorrespondingauthor{Junwu Xiong}{xiongjunwu.1@jd.com}

  \icmlkeywords{Machine Learning, ICML}

  \vskip 0.3in
]



\printAffiliationsAndNotice{\icmlEqualContribution}

\begin{abstract}
 In video generation models, particularly world models, training large-scale video diffusion Transformers (such as DiT and MMDiT) poses significant computational challenges due to the extreme variance in sequence lengths within mixed-mode datasets. Existing bucket-based data loading strategies typically rely on "equal token length" constraints. This approach fails to account for the quadratic complexity of self-attention mechanisms, leading to severe load imbalance and underutilization of GPU resources. 
  This paper proposes \textit{AdaptiveLoad}, an integrated optimization framework consisting of two core components: 
(1) A dual-constraint adaptive load balancing system, which eliminates long-sequence bottlenecks by simultaneously limiting memory consumption and computational load ($B \times S^p \le M_{\text{comp}}$); 
(2) A fused LayerNorm-Modulate CUDA kernel, which utilizes a D-tile coalesced reduction strategy to increase throughput and alleviate memory pressure. 
Experimental results on the Wan 2.1 world model demonstrate that our method reduces the computational imbalance rate from 39\% to 18.9\%, improves peak VRAM utilization efficiency by 22.7\%, and achieves an overall training throughput increase of 27.2\%.
\end{abstract}

\section{Introduction}
\label{sec:intro}

The paradigm of generative artificial intelligence is undergoing a profound transformation from static image synthesis to high-fidelity dynamic video generation, an evolution primarily driven by the exceptional scalability of Diffusion Transformer (DiTs) architectures~\cite{peebles2023scalable}. Unlike image models that process fixed resolutions, modern video generators~\cite{skorokhodov2022stylegan,khachatryan2023text2video} must support highly heterogeneous data shapes, ranging from single-frame static images to long-duration, high-frame-rate complex video sequences. This diversity in data dimensionality introduces severe \textit{Bucket Management} challenges. In standard distributed training frameworks, the synchronization of model parameters relies on the \texttt{AllReduce}~\cite{patarasuk2009bandwidth} collective communication operation at the end of each step. Consequently, the global computational efficiency is strictly constrained by the ``Long-tail Effect''~\cite{constantinides2021rigorous, guan2026rl}, where the global step latency is stochastically determined by the slowest GPU in the cluster: 
\begin{equation}
    T_{\text{sync}} = \max_{i \in \{1, \dots, N\}} \{T_i\}
\end{equation}
where $T_i$ denotes the execution time of the $i$-th GPU.

However, the ``Equal-token'' bucket strategy~\cite{diao2025nemotron} commonly adopted in current industrial pipelines severely overlooks the non-linear computational characteristics of the core self-attention mechanism. While maintaining a constant token count per batch provides a first-order approximation of memory usage, it fails to account for the quadratic complexity scaling of Transformers. Experimental observations in our study indicate that the correlation between training step latency and the total number of tokens is merely $R \approx 0.35$, whereas it exhibits an extremely strong correlation ($R \approx 0.92$) with the square of the sequence length. This systematic bias in load estimation causes long-sequence buckets to become significant ``computational stragglers''~\cite{yan2026scalabletrainingmixtureofexpertsmodels}. This imbalance forces GPUs processing long-sequence data into a state of chronic computational overload, while other nodes face idle wait times (synchronization bubbles) of several seconds, resulting in a substantial waste of high-value computing power in large-scale GPU clusters. This gap between linear token counting and quadratic computational reality motivates our first core contribution: a dual-constraint scheduling logic that reconciles memory safety with computational parity.

Parallel to the scheduling bottleneck, the architectural evolution towards Multimodal DiT (MMDiT) modules~\cite{reuss2024multimodal}---exemplified by SD3~\cite{esser2024scaling} and Wan 2.1~\cite{wan2025wan}---has introduced new low-level performance hurdles. In these architectures, Adaptive Layer Normalization (AdaLN) serves as the critical bridge connecting timestep embeddings with multimodal feature flows~\cite{zhang2024dim}, being invoked hundreds of times per training iteration~\cite{team2025kling}. Although memory-efficient attention kernels like FlashAttention~\cite{dao2022flashattention, dao2023flashattention} have significantly mitigated the core attention bottleneck, peripheral auxiliary operators have emerged as the new primary performance ceiling due to the ``Memory Wall''~\cite{wulf1995hitting} problem. 

Standard AdaLN implementations typically rely on a sequence of discrete CUDA kernel calls. In the forward pass, this leads to frequent, redundant round-trips of intermediate tensors between the GPU's Streaming Multiprocessors (SMs) and the High Bandwidth Memory (HBM). This memory-bound behavior is further exacerbated during backpropagation. The gradient reduction operations for modulation parameters (scale and shift) are hindered by inefficient \textit{strided memory access} patterns. Because the hidden dimension $D$ is the contiguous dimension in memory, accumulating gradients across the sequence dimension $N$ fails to trigger hardware-level \textit{coalesced access}~\cite{harris2007optimizing}. This leads to effective memory bandwidth utilization far below the theoretical peak of modern HBM3/HBM3e systems. The fundamental inefficiency of these ``auxiliary'' operators in the MMDiT backbone necessitates a hardware-aware redesign, shifting the paradigm from discrete operator execution to fused, bandwidth-optimized kernels.

\begin{figure}
    \centering
    \includegraphics[width=0.8\linewidth]{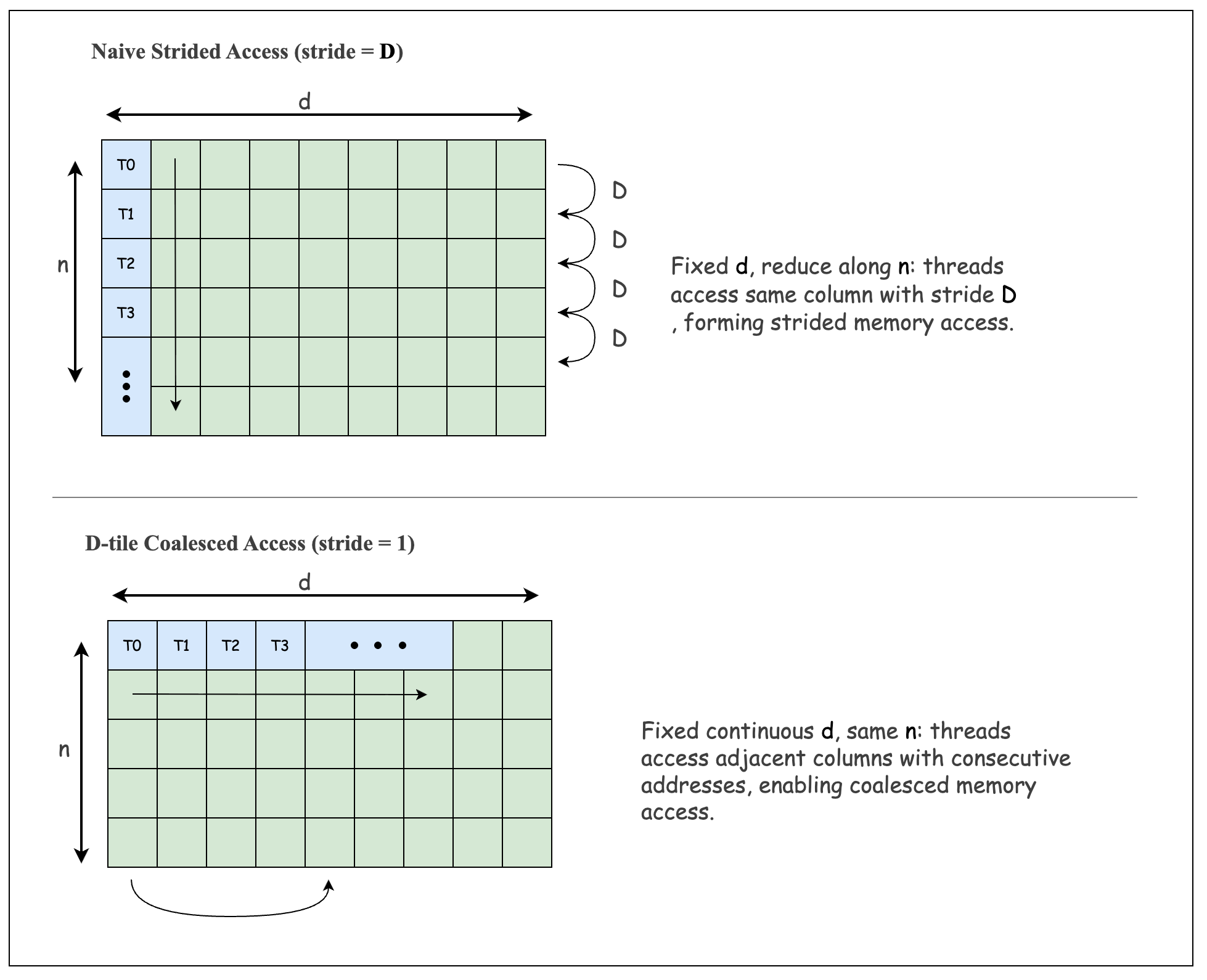}
    \caption{Comparison of Naive Access and D-tile Coalesced Access Patterns}
    \label{fig:dtile-access}
\end{figure}

To address these multi-level challenges, this paper proposes a full-stack optimization framework named \textit{\textbf{AdaptiveLoad}}. Our approach synergizes data-driven load balancing with hardware-aware operator fusion to reclaim lost training efficiency. First, we transcend the traditional linear load assumption by introducing a dual-constraint adaptive load balancing strategy. This mechanism dynamically optimizes the batch size of each bucket by simultaneously enforcing a linear memory capacity boundary and a polynomial computational complexity boundary. Second, we design an automated shape benchmarking and cost-fitting system. By capturing execution traces in a live distributed environment and utilizing non-linear regression, the system achieves precise modeling of computational costs, effectively eliminating the sub-optimality inherent in manual empirical tuning. Finally, addressing the AdaLN efficiency gap, we developed a Fused AdaLN CUDA kernel and an innovative D-tile coalesced reduction strategy, as show in Figure~\ref{fig:dtile-access}. This innovation fundamentally restructures memory access patterns---transforming strided access into coalesced HBM transactions---thereby significantly enhancing the arithmetic intensity and maximizing effective bandwidth utilization during gradient reduction.

The main contributions of this paper are summarized as follows:
\begin{itemize}
    \item We provide a rigorous quantification of the load mismatch phenomenon in video diffusion training and propose a dual-constraint adaptive bucket scheduling algorithm. This approach effectively eliminates the synchronization stragglers caused by long-sequence quadratic complexity, reducing the load imbalance rate by up to 50\%.
    \item We design and implement an end-to-end training bottleneck quantization and feedback closed-loop system. This shifts distributed training configurations from being heuristic-driven to data-driven, utilizing automated cost-model fitting via our \textit{Shape Benchmark} to optimize cluster-wide resource utilization.
    \item We develop a high-performance Fused AdaLN CUDA kernel specifically optimized for MMDiT backbones. By introducing \textbf{D-tile coalesced reduction}, we resolve the long-standing strided access bottleneck in backpropagation, increasing training throughput by over 25\% while simultaneously reducing the activation memory footprint, thereby providing a robust foundation for scaling ultra-long sequence video generation models.
\end{itemize}

\section{Related Work}

\subsection{Memory Optimization Techniques}
In large-scale model training, memory capacity remains the core bottleneck restricting the scaling of sequence lengths. Existing optimization research primarily focuses on three dimensions: first, memory-efficient attention operators (such as the FlashAttention series~\cite{dao2022flashattention,dao2023flashattention,shah2024flashattention}) effectively reduce the spatial complexity of self-attention from $O(S^2)$ to $O(S)$ through tiling and operator restructuring; second, gradient and parameter sharding techniques (such as the ZeRO series~\cite{rajbhandari2020zero}) eliminate redundant model state occupancy by leveraging distributed storage; and finally, activation checkpointing and offloading strategies (such as DeepSpeed-Ulysses~\cite{jacobs2023deepspeed}) alleviate peak memory pressure by trading computation for space or utilizing CPU memory as a secondary cache. However, most of these general-purpose techniques are designed for fixed-dimension sequences and fail to fully account for the dynamic shape characteristics of multimodal data in video generation models. Particularly in complex scenarios involving mixed training of video and images, fixed-stride optimization strategies struggle to cope with drastic fluctuations in memory demand, leading to constrained computational efficiency for long-sequence samples~\cite{han2026making,liu2025elasticmm}.

\subsection{Load Balancing in Distributed Training}
Load balancing in distributed deep learning has traditionally focused on uniform batch processing in data parallelism. For Transformer architectures, the industry commonly adopts a "constant token budget" heuristic strategy, which predicts computational load by constraining $B \times S$ to a constant. However, as sequence lengths $S$ in video generation tasks scale to the order of $10^5$, the quadratic complexity $O(S^2)$ of the self-attention mechanism renders this linear-assumption-based heuristic completely ineffective. While existing research such as Megatron-LM~\cite{shoeybi2019megatron} proposes sophisticated 3D parallelism partitioning, or addresses data heterogeneity through dynamic task scheduling (e.g., ByteScale~\cite{ge2025bytescale}), these solutions mostly focus on a single linguistic modality. When dealing with world models like Wan 2.1~\cite{wan2025wan}, the significant differences in computational patterns between multi-modal encoding and language modeling tasks prevent existing schedulers from effectively resolving load mismatches between modalities, resulting in severe GPU idle waste at global synchronization points (Barriers). To mitigate this, ByteScale~\cite{ge2025bytescale} introduces a flexible framework to handle the mismatch between data heterogeneity~\cite{huang2024distmm,zhang2025disttrain} and static mesh. Similarly, MegaScale-Data~\cite{zhao2026megascaledatascalingdataloadermultisource} addresses the workload imbalance specifically in multisource foundation model training by optimizing the distributed dataloader architecture.

\subsection{Hardware-Aware Operator Fusion}
As a key technology to break the "Memory Wall"~\cite{gholami2024ai} bottleneck, operator fusion significantly enhances arithmetic intensity by reducing redundant reads and writes of intermediate results in the GPU's HBM. Representative works include FlashAttention-2~\cite{dao2023flashattention} for attention mechanisms and FusedSoftmaxCrossEntropy~\cite{ge2025bytescale} for loss functions. Although these solutions have achieved success in core operators, the overhead of "auxiliary" operators within actual Transformer Blocks---such as LayerNorm---is relatively increasing as attention efficiency improves, gradually becoming a new performance bottleneck. Existing fusion schemes typically overlook the specific strided memory access patterns inherent in AdaLN during backpropagation. This non-contiguous access behavior severely hinders the GPU's coalescing properties; especially when processing ultra-long sequences, low bandwidth utilization in reduction operations has become a critical obstacle preventing further increases in training throughput.

\section{AdaptiveLoad: Dual-Constraint Load Balancing and Fused Modulated Kernels}

\begin{figure*}[h!]
    \centering
    \includegraphics[width=0.7\linewidth]{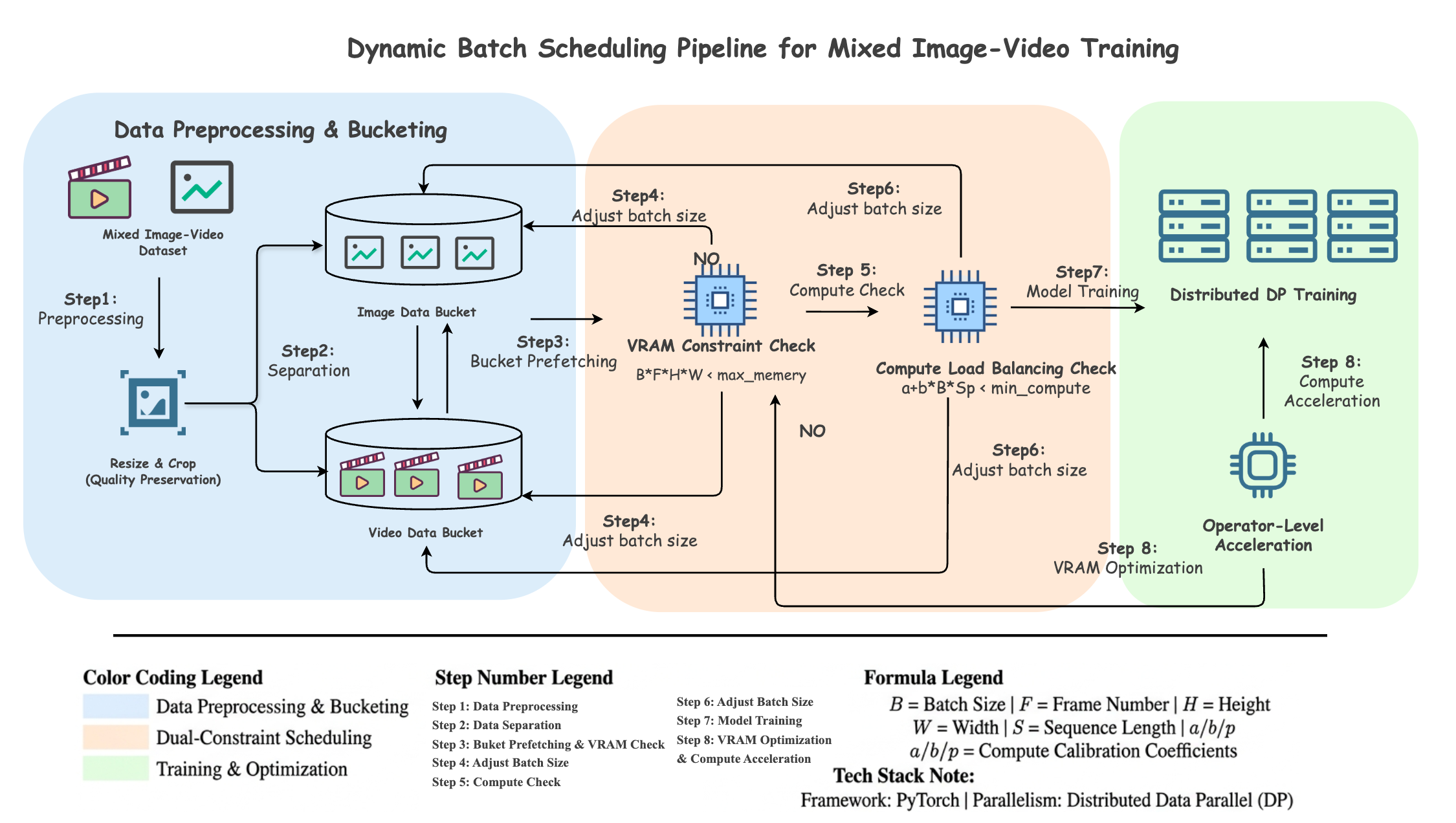}
    \caption{The data pipeline for joint image-video training with adaptive bucketing.
    }
    \label{fig:framework}
\end{figure*}

\subsection{Overview}
The training framework proposed in this paper addresses the dual challenges of computational heterogeneity and memory constraints inherent in large-scale video diffusion Transformers (e.g., DiT, MMDiT), as shown in Figure~\ref{fig:framework}. We have constructed a full-stack optimization system spanning from high-level distributed scheduling to low-level kernel execution.

The core of this approach is the Dual-Constraint Adaptive Load Balancing strategy, which moves away from traditional fixed-token allocation in favor of an empirical complexity power-law distribution. This mitigates the "long-tail" latency issues caused by $O(S^2)$ complexity in long-sequence video data buckets. To ensure this strategy aligns with actual hardware performance, we employ a data-driven Shape Benchmarking workflow that fits a parameterized overhead model with real-world synchronization telemetry, enabling precise calibration of computational boundaries.

Building upon this macro-system equilibrium, we further implement \textit{Fused AdaLN CUDA kernels} to optimize the micro-architectural execution logic of the MMDiT backbone. By fusing normalization and modulation operations and introducing a \textit{D-tile Coalesced Reduction} strategy for gradient computation, the system significantly alleviates HBM bottlenecks and eliminates redundant memory access. Together, these components form a closed-loop optimization system that dynamically adapts to diverse input morphologies while maintaining near-peak hardware utilization across distributed GPU clusters.

\subsection{Dual-Constraint Adaptive Load Balancing \& Cost Fitting}
To address the computational synchronization bottlenecks faced by video diffusion Transformers when processing heterogeneous data, we implement a dual-boundary strategy to determine the batch size $B$ for any given bucket shape defined by sequence length $S$. Traditional "equal-token" allocation schemes (which constrain $B \times S = \text{Constant}$) overlook the $O(S^2)$ computational complexity of the core Transformer attention operators, causing long-sequence buckets to become significant "long-tail" bottlenecks in distributed environments. Experimental observations show that the correlation between training step latency and total token count is only $0.35$, whereas the correlation with $B \times S^p$ reaches as high as $0.92$.

To eliminate this imbalance, the system first calculates the logical sequence length $S = S_{\text{text}} + S_{\text{visual}}$ for each data shape $(n_{\text{frame}}, H, W)$ after VAE encoding, where $S_{\text{visual}}$ is compressed according to temporal and spatial downsampling factors (8 and 16, respectively). Subsequently, we simultaneously apply a linear memory constraint and a polynomial computational constraint to each bucket, taking their intersection as the final batch size $B_{\text{shape}}$:
\begin{equation}
\label{equ:shape}
B_{\text{shape}} = \max(1, \min(\lfloor M_{\text{mem}} / S \rfloor, \lfloor M_{\text{comp}} / S^p \rfloor))
\end{equation}
where $M_{\text{mem}}$ represents the upper memory bound determined by GPU capacity and static model overhead, $M_{\text{comp}}$ is the upper bound for computational load, and $p$ denotes the empirical exponent of attention complexity. Under this strategy, short-sequence buckets are governed by memory limits to maintain high throughput, while long-sequence buckets trigger the computational constraint. By actively reducing the batch size, we ensure that GPUs holding long-sequence data do not slow down the global synchronization time $\text{step\_time\_sync}$ due to $O(S^2)$ load surges.

To precisely quantify the computational constraint parameter $M_{\text{comp}}$ and the exponent $p$, the system incorporates a data-driven shape benchmarking system. This system operates within a real distributed environment (e.g., FSDP communication paths) and measures the mapping matrix $(B, S) \to \text{step\_time\_sync}$ via synthetic pixel scans that exclude data-loading I/O jitter. To balance measurement accuracy and cost, the system employs a Throughput Sweep mode, prioritizing multi-level batch size tests for long-sequence buckets where $S \ge 20,000$ to capture performance characteristics in the compute-intensive regime.

Based on the collected benchmark data, we construct a parameterized cost model: $\text{step\_time\_sync} \approx a + b \times B \times S^p$. By performing a grid search for optimization within the interval $p \in [1.6, 2.4]$, we select the $\hat{p}$ that maximizes the coefficient of determination $R^2$ as the model parameter. Subsequently, based on a preset training target step latency $\text{target\_sync}$, we back-derive the computational load upper bound $M_{\text{comp}} = (\text{target\_sync} - a) / b$. This process establishes a closed-loop optimization framework: it monitors the waiting time $\text{wait\_sync}$ of each GPU in real-time, identifies the primary bottleneck using bottleneck analysis tools, and dynamically recalibrates bucket configurations. 


\subsection{Fused LayerNorm-Modulate CUDA Kernels}

\begin{figure}
    \centering
    \includegraphics[width=0.8\linewidth]{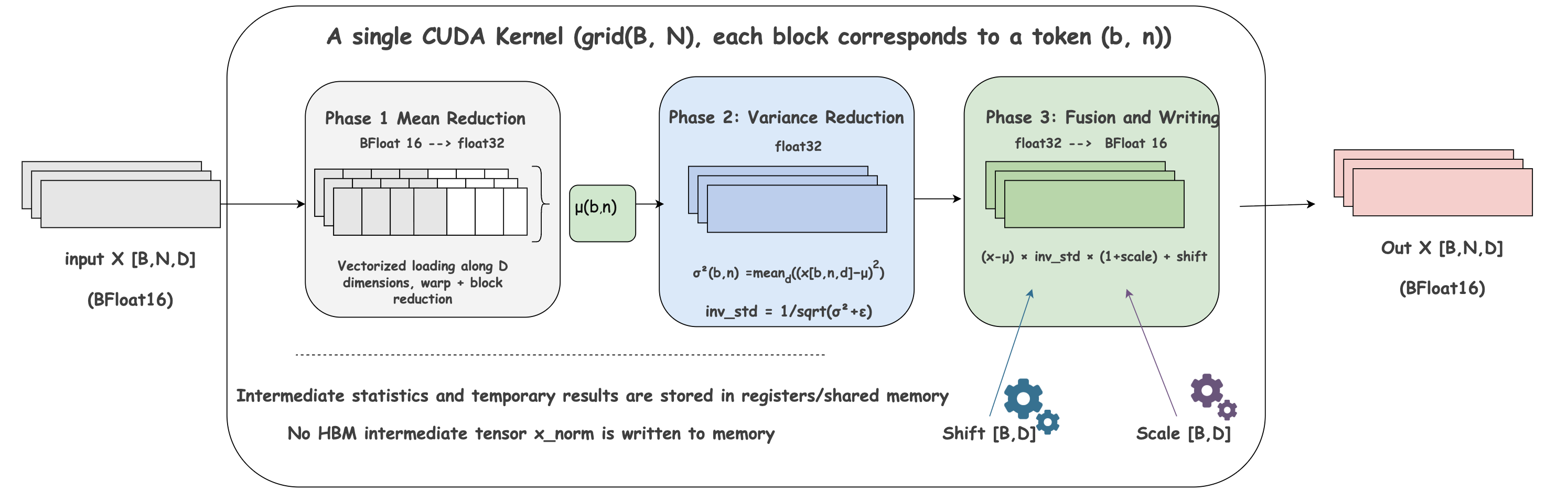}
    \caption{Schematic diagram of forward fusion kernel data flow}
    \label{fig:forward_fusion}
\end{figure}
In the MMDiT architecture, AdaLN is invoked with high frequency. Traditional decoupled operator implementations introduce substantial bandwidth waste and scheduling overhead due to frequent HBM reads/writes of intermediate tensors (such as the normalized $x_{\text{norm}}$). This paper designs and implements a fully fused CUDA kernel aimed at completing the entire normalization and modulation process in a single kernel launch. In the forward pass, the kernel uses the Token as the unit of parallelism, utilizing registers and Shared Memory to temporarily store intermediate statistics, as show in Figure~\ref{fig:forward_fusion}. Mean $\mu$ and variance $\sigma^2$ are calculated via two-stage reduction (Warp-level Shuffle and Block-level Reduction), and linear transformations are written out directly in combination with modulation parameters during a third traversal of dimension $D$. This eliminates approximately $131$ GB/step of redundant HBM access in a 40-layer MMDiT model, significantly alleviating memory pressure.

\begin{figure}
    \centering
    \includegraphics[width=0.8\linewidth]{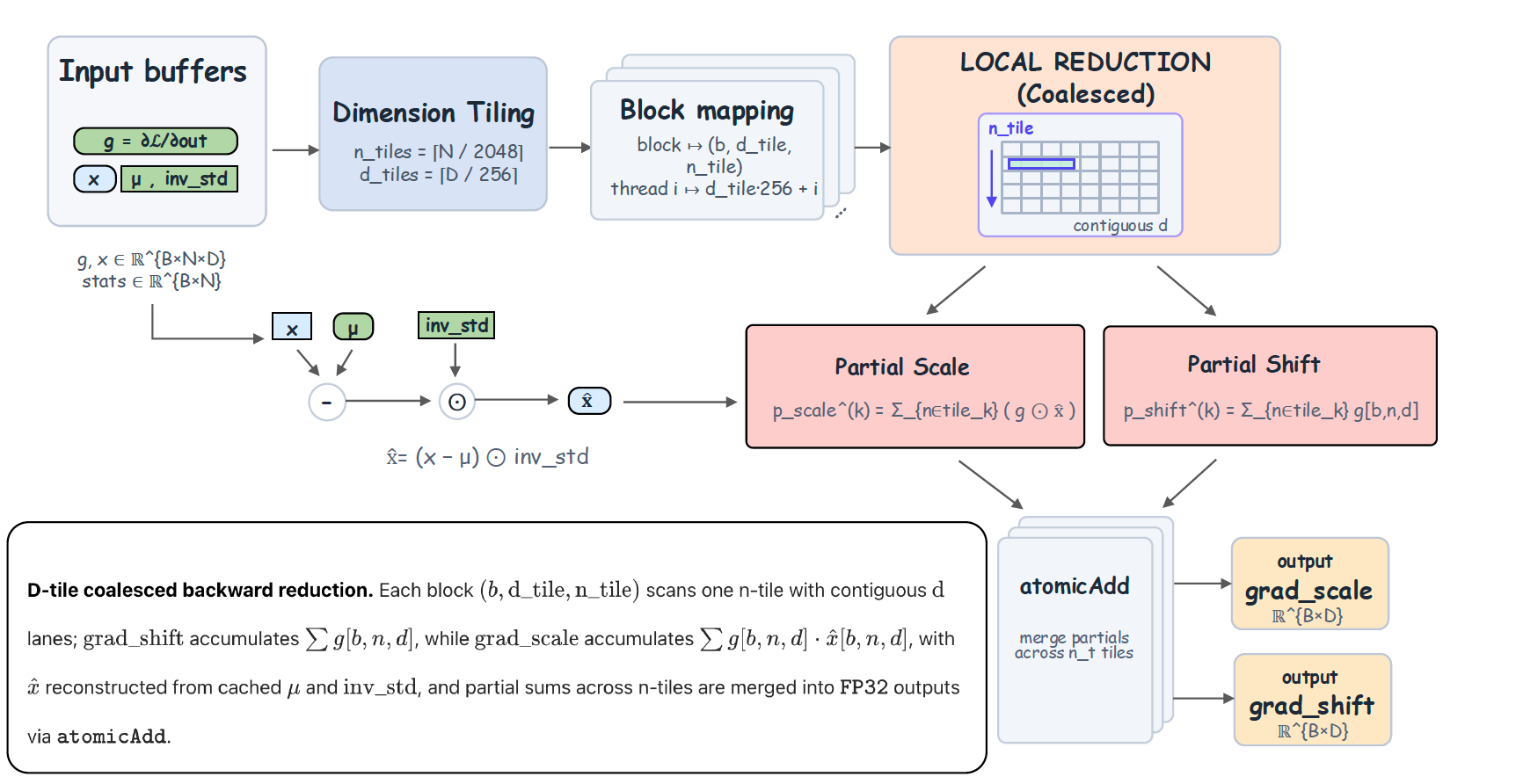}
    \caption{Schematic diagram of grad\_shift/grad\_scale reduction process. 
    }
    \label{fig:grad_reduction}
\end{figure}

In the backpropagation stage, specifically for the computation of modulation parameter gradients $\nabla_{\text{shift}}$ and $\nabla_{\text{scale}}$, we propose an innovative \textbf{D-tile Coalesced Reduction} strategy.
Traditional gradient reduction occurs along the sequence dimension $N$. Since the feature dimension $D$ is the contiguous storage dimension, this results in strided, non-contiguous memory access, which severely limits memory bandwidth. The D-tile strategy swaps the loop hierarchy, partitioning the Grid into a $(d_{\text{tile}}, n_{\text{tile}})$ layout. Each thread is fixed to a feature index $d$ and performs vertical accumulation along the sequence tiles, as show in Figure~\ref{fig:grad_reduction}. In this mode, the physical addresses accessed by the 32 threads within the same Warp are perfectly contiguous, thereby boosting memory bandwidth utilization to near-theoretical peaks. Furthermore, the kernel caches computed statistics in global memory for subsequent reuse, avoiding redundant reduction calculations and ensuring natural compatibility and numerical stability in distributed sequence-parallel scenarios.


\subsection{Activation Computational Graph Simplification and Memory Recovery}
From the perspective of the computational graph, operator fusion not only reduces hardware-level memory access overhead but also achieves significant simplification of the graph within the deep learning framework's automatic differentiation (Autograd) engine. In traditional discrete operator implementations, each constituent operation (e.g., \texttt{Mean}, \texttt{Var}, \texttt{Standardize}, \texttt{Mul}, \texttt{Add}) generates an independent node in the computational graph. This forces the system to retain the forward outputs of each node as "activations" to facilitate gradient computation during the backward pass. Consequently, the activation footprint scales linearly with the length of the operator chain.

By integrating the entire path of AdaLN into a single fused CUDA kernel, we collapse the computational subgraph—originally consisting of 5--8 discrete nodes—into a single atomic node. Under this paradigm, only the final output of the kernel needs to be stored in the activation checkpoint. All intermediate statistics, including means, variances, and intermediate scaling factors, complete their entire lifecycle within registers instantaneously, bypassing the framework's global memory management system. This redundancy elimination at the computational graph level provides a theoretical "negative entropy" for video generation models, serving as critical infrastructure for supporting ultra-long sequence lengths (\textit{Seqlen}).

\section{Experimental Evaluation}

\subsection{Experimental Setup}
\label{subsec:setup}
We adopt \textbf{Wan 2.1}~\cite{wan2025wan} as our primary architecture, implemented within the \textbf{DiffSynth-Studio}~\cite{modelscope2024diffsynthstudio} ecosystem to ensure compatibility with large-scale generative benchmarks~\cite{fu2025polyart, qiu2026emovid, yang2025towards, he2025diffthinker}. The system is stress-tested using a mixed corpus of 10 million samples from \textbf{WebDataset}~\cite{huggingface2025webdataset} and \textbf{Koala-36m}~\cite{wang2025koala}, creating extreme sequence length variance. 
To evaluate efficiency, we employ three metrics: \textbf{Throughput Efficiency} ($\Theta = \frac{B \times [ \frac{F-1}{\lambda} + 1 ] \times \frac{W}{\gamma} \times \frac{H}{\eta}}{T}$), reflecting latent units processed per unit time; \textbf{Load Balancing Efficiency} ($CV_{\text{step}} = \frac{\text{len}_{\text{max}} - \text{len}_{\text{min}}}{\text{len}_{\text{max}}}$), measuring synchronization overhead; and \textbf{Physical Load Pressure} ($O = B \times S^2$), which models the quadratic complexity of self-attention.

\subsection{Overall System Throughput and Scalability}

The experimental results provide empirical evidence for the significant superiority of our integrated optimization strategy in enhancing system throughput. By mitigating computational imbalances and kernel-level redundancies, we achieve substantial gains across different cluster scales.

\begin{figure*}[t]
    \centering
    \begin{subfigure}{0.48\textwidth}
        \centering
        \includegraphics[width=\linewidth]{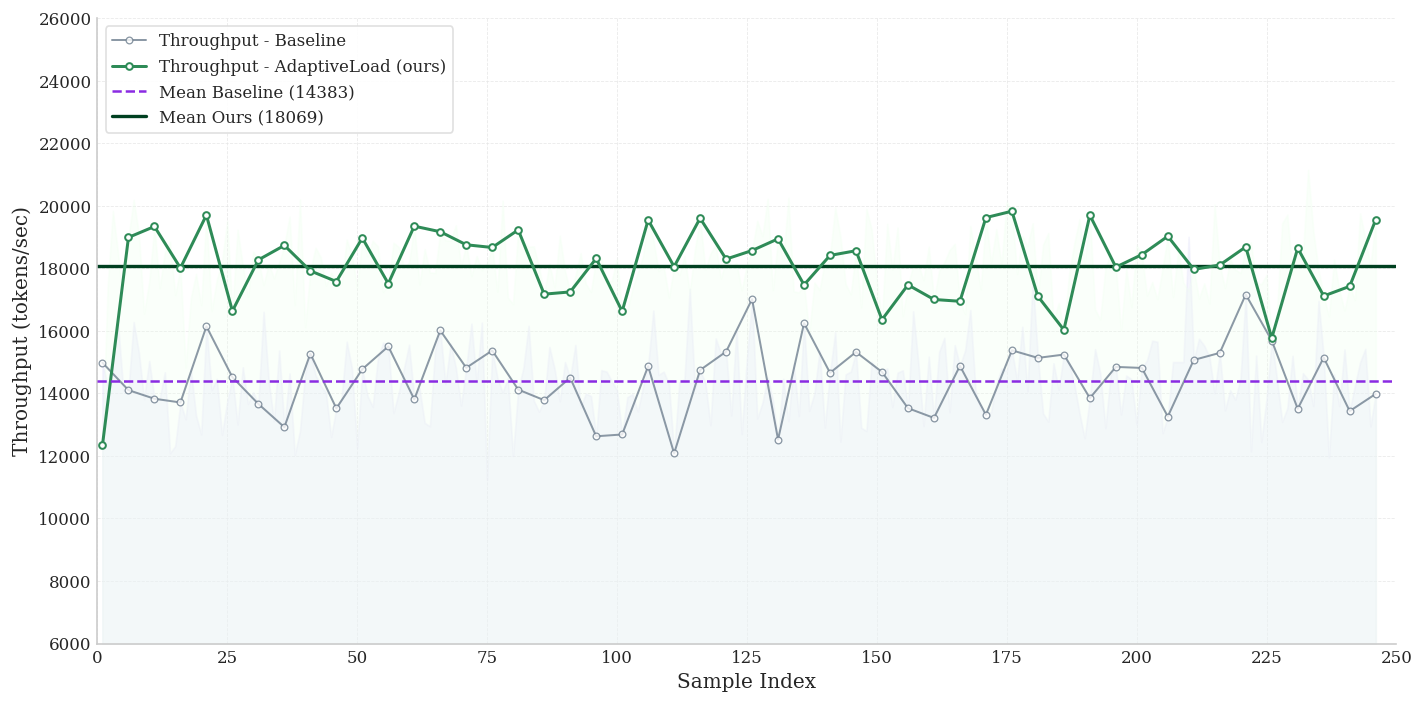}
        \caption{8-GPU Configuration}
        \label{fig:throughput_8}
    \end{subfigure}
    \hfill
    \begin{subfigure}{0.48\textwidth}
        \centering
        \includegraphics[width=\linewidth]{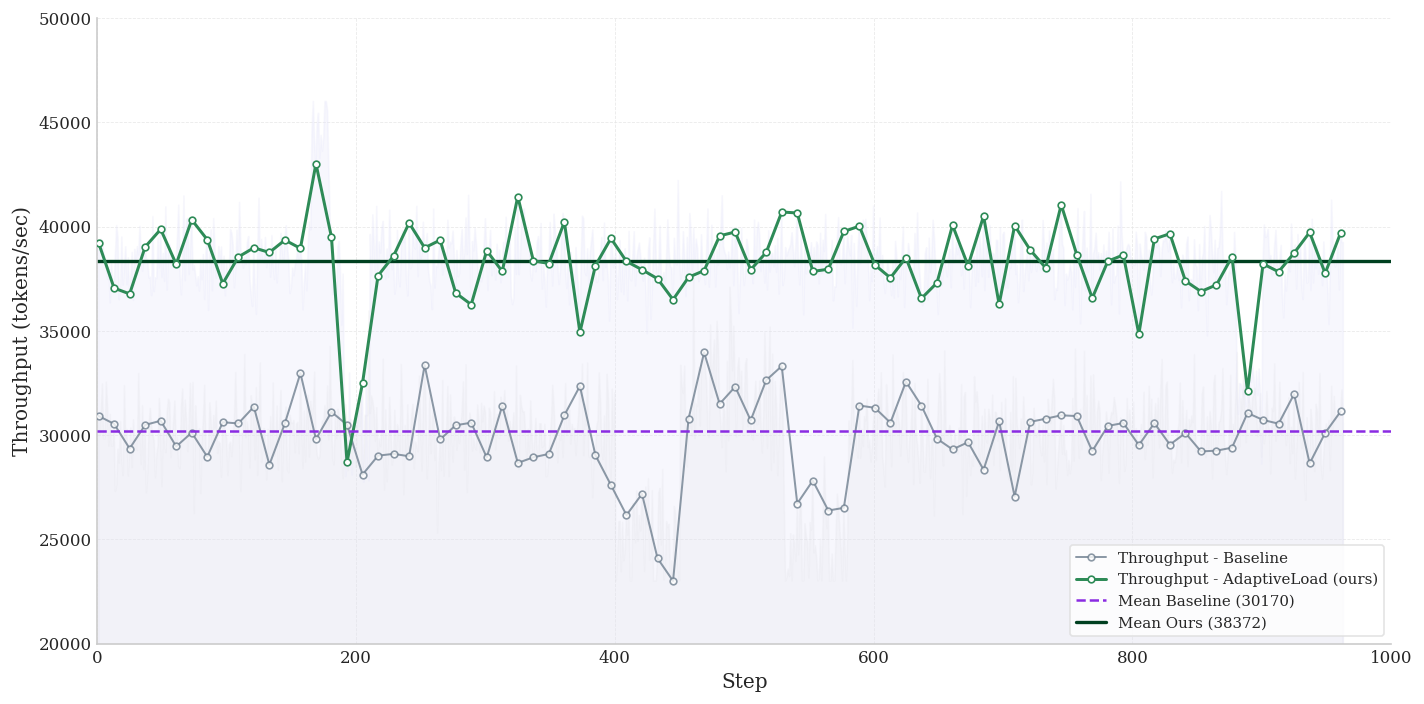}
        \caption{16-GPU Configuration}
        \label{fig:throughput_16}
    \end{subfigure}
    \caption{Throughput (tokens/sec) comparison between Baseline and AdaptiveLoad. The shaded areas represent the raw fluctuations per step, while the solid lines indicate the moving average.}
\end{figure*}

\textbf{Throughput Gains in 8-GPU Cluster:} 
As illustrated in Figure~\ref{fig:throughput_8}, the initial introduction of operator fusion techniques (e.g., \textit{Fused AdaLN}) effectively reduced redundant HBM accesses. However, the most significant leap occurred after deploying the dual-constraint adaptive load balancing strategy. The mean training throughput surged from 14,383 tokens/sec to 18,069 tokens/sec, representing a substantial efficiency gain of 25.6\%. Notably, the AdaptiveLoad curve (dark green) consistently maintains a higher floor compared to the Baseline (grey), indicating that even in worst-case sequence distributions, our system preserves high hardware utilization.

\textbf{Scaling Advantage in 16-GPU Cluster:} 
This improvement is further amplified in the 16-GPU scaling experiments (Figure~\ref{fig:throughput_16}). The average throughput rose from 30,170 tokens/sec to 38,372 tokens/sec, achieving an optimization margin of 27.18\%. The widening gap between the two configurations as the cluster scale doubles highlights a critical insight: as parallel scale expands, the synchronization overhead caused by the ``long-tail effect'' of heterogeneous sequences becomes the primary bottleneck. 

\textbf{Analysis of the ``Long-Tail'' Suppression:} 
The Baseline throughput exhibits frequent downward spikes (notably around Step 200 and Step 450 in Figure~\ref{fig:throughput_16}), where massive computational imbalances force high-performance nodes into prolonged idle states. AdaptiveLoad successfully flattens these throughput dips. By utilizing precise cost modeling and dynamic bucketing, our method liberates the latent computing power previously restricted by synchronization barriers. This robust scalability demonstrates that AdaptiveLoad is particularly well-suited for large-scale training on heterogeneous video datasets, where sequence length variance is inherently high.
\subsection{Multidimensional Load Balancing Evaluation}
To analyze load balancing in distributed training, we performed deep quantification across two dimensions: load balancing efficiency and computational complexity.


\subsubsection{load balancing efficiency}

\begin{figure}
    \centering
    \includegraphics[width=1\linewidth]{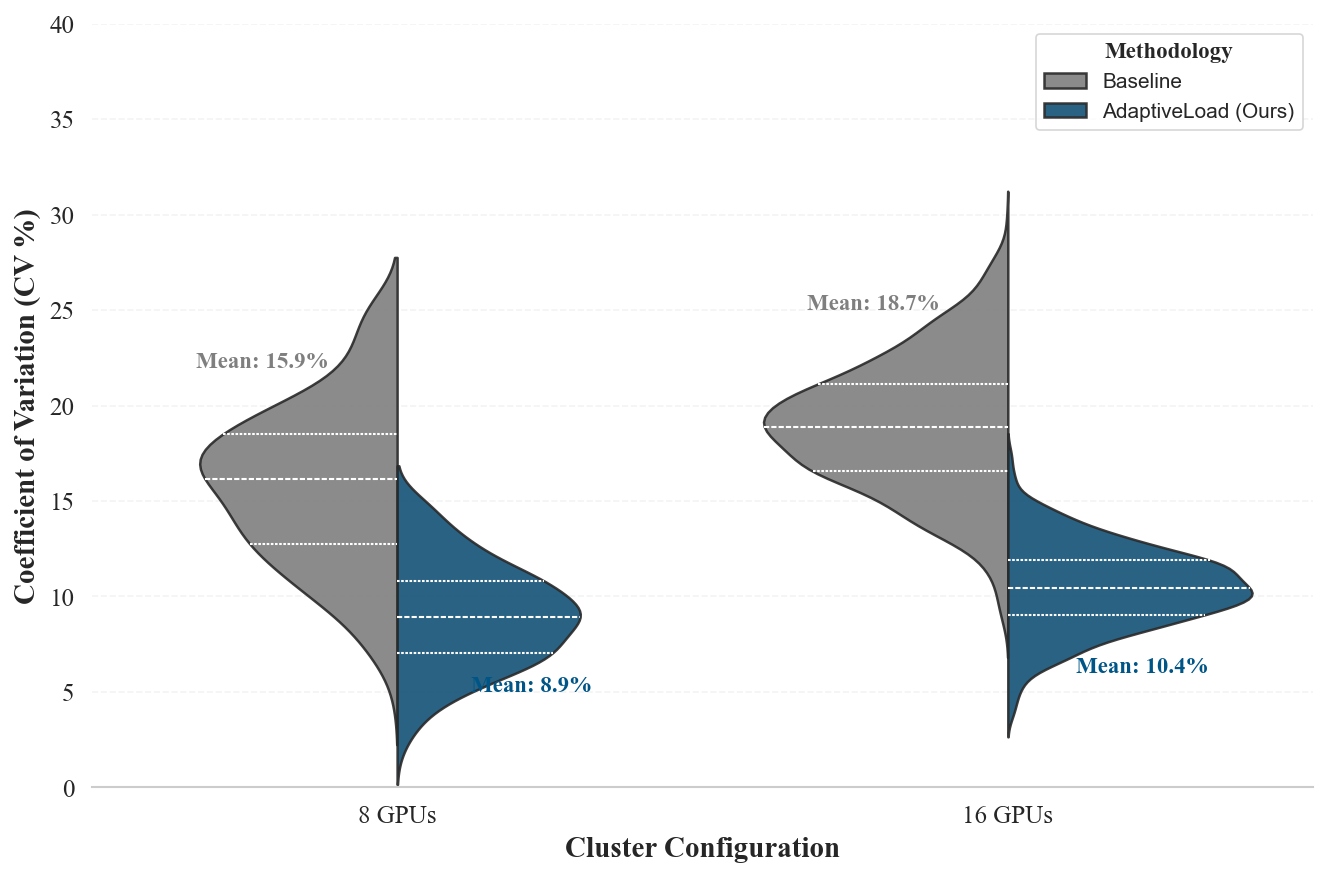}
    \caption{Comparison of $CV_{\text{step}}$ distribution between Baseline and AdaptiveLoad across 8-GPU and 16-GPU configurations.}
    \label{fig:CV1_compare}
\end{figure}

As illustrated in Figure~\ref{fig:CV1_compare}, AdaptiveLoad demonstrates a superior capability in harmonizing cross-GPU workloads. In the 8-GPU configuration, the mean $CV_{\text{step}}$ drops sharply from 15.9\% to 8.9\%, with the distribution exhibiting a significantly narrower body. This shift indicates a transition from a high-variance, unpredictable scheduling state to a tightly regulated, load-balanced regime where the majority of iterations operate under highly synchronized conditions.

The optimization effect becomes more pronounced as the cluster scales. In the 16-GPU configuration, while the Baseline mean $CV_{\text{step}}$ rises to 18.7\% due to increased fragmentation, AdaptiveLoad effectively suppresses this trend, maintaining a robust mean of 10.4\%. This improvement is attributed to the dynamic bucketing mechanism, which intelligently re-aligns input dimensions in real-time to mitigate the ``straggler effect,'' ensuring near-uniform peak memory usage and maximizing effective hardware throughput.

\subsubsection{Computational Complexity and Load Alignment}

\begin{figure}
    \centering
    \includegraphics[width=1\linewidth]{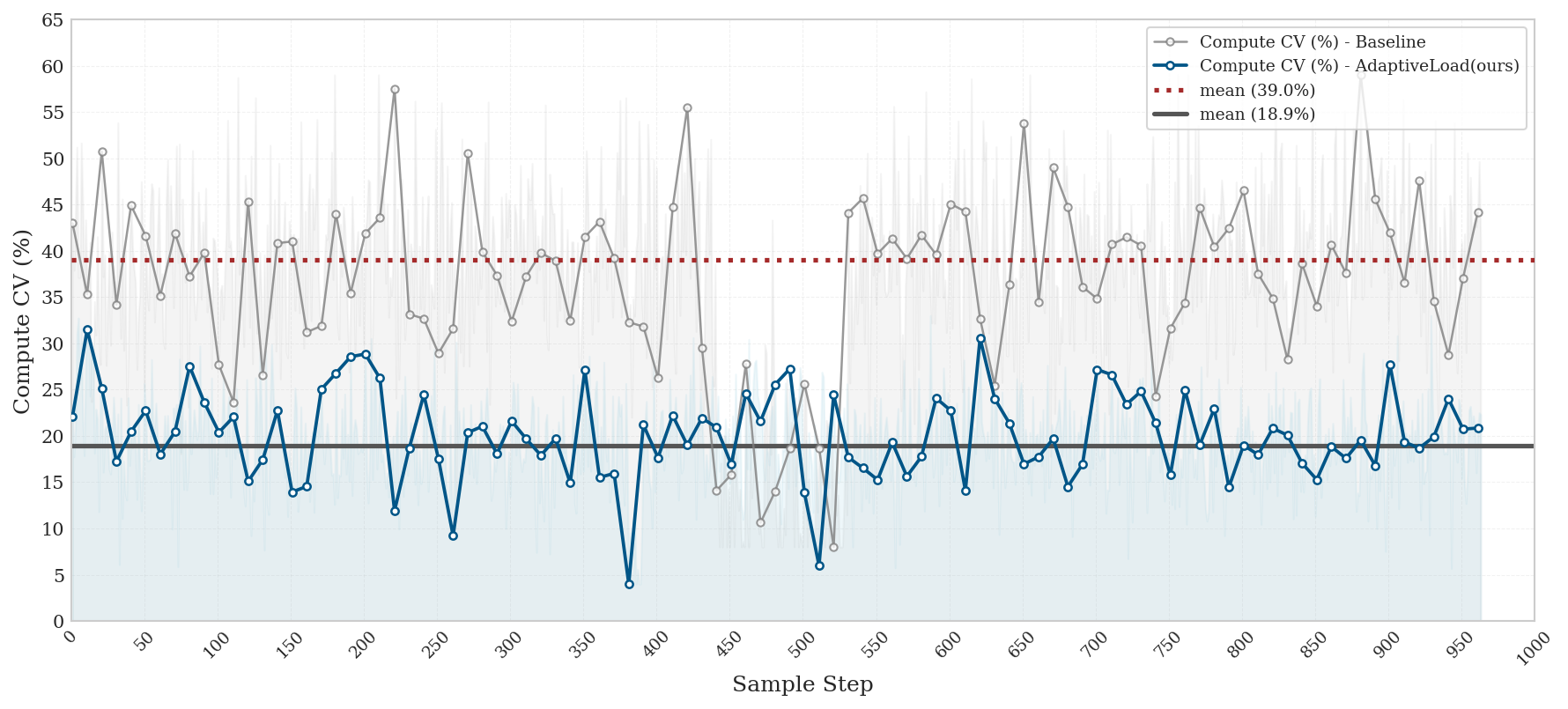}
    \caption{Comparison of Compute CV (\%) between Baseline and AdaptiveLoad. The metric reflects the variance in quadratic computational pressure ($B \times S^2$) across 16 GPUs.}
    \label{fig:cv2_compare}
\end{figure}

The empirical results in Figure~\ref{fig:cv2_compare} reveal a substantial optimization in hardware utilization. In a high-pressure 16-GPU environment, the variation coefficient of computational complexity plummeted from a baseline mean of \textbf{39.0\%} to \textbf{18.9\%} with AdaptiveLoad. This reduction is critical for several reasons:

\begin{itemize}
    \item \textbf{Mitigating Quadratic Imbalance:} As shown by the grey curve in Figure~\ref{fig:cv2_compare}, the Baseline frequently exhibits extreme spikes in Compute CV (exceeding 55\%), where the quadratic nature of attention $S^2$ exacerbates even minor differences in sequence distribution. AdaptiveLoad (blue curve) successfully flattens these peaks by actively perceiving the $S^p$ weight of each bucket and executing a refined batch size reduction strategy for long-sequence clusters.
    
    \item \textbf{Eliminating Synchronization Bubbles:} The drastic narrowing of the CV gap directly translates to minimized load deviation at synchronization points (Barriers). By ensuring that each GPU processes a similar total $O$ per step, AdaptiveLoad effectively fills the \textit{idle bubbles} that traditionally plague distributed training. This transformation converts previously wasted ``sync-waiting periods'' into effective ``floating-point operation periods.''
    
    \item \textbf{Deterministic Throughput:} Unlike the Baseline, which fluctuates wildly across Sample Steps, AdaptiveLoad maintains a consistently low Compute CV. This stability ensures predictable iteration times and prevents thermal throttling or power surges often caused by highly imbalanced, bursty computational workloads.
\end{itemize}

Through this active weight-aware scheduling, AdaptiveLoad moves beyond simple padding-minimization to achieve true hardware-level load symmetry.

\subsection{Evaluation of Operator Fusion and Memory Efficiency}
\label{subsec:operator_fusion}

To further validate the superior performance of operator fusion strategies in mitigating ``Memory Wall'' bottlenecks, we implemented deep kernel fusion optimizations for critical operator chains within the MMDiT architecture. This experimental suite focuses on core forward and backward propagation operator combinations, including Adaptive Layer Normalization and Modulation (\texttt{Norm + Modulate}), Query/Key Normalization (\texttt{Q-Norm + K-Norm}), and Gated Normalization (\texttt{Gate + Norm}). By consolidating these discrete logics into unified CUDA kernels, we aim to fundamentally eliminate the high-frequency round-trips of intermediate activations to High Bandwidth Memory (HBM), thereby increasing computational density while reclaiming precious memory resources for long-sequence training.

\begin{table}[htbp]
\centering
\caption{System-level Comparison of Performance and Memory Metrics Before and After Operator Fusion}
\label{tab:fuson_compare}
\small 
\begin{tabular}{lccc} 
\toprule
\textbf{Metric} & \textbf{Baseline} & \textbf{Fused} & \textbf{Fused} \\ 
\textbf{Metric} & \textbf{(Max)} & \textbf{(same)} & \textbf{(Max)} \\ 
\midrule
Allocated Mem (GB) & 88     & 87.9                  & 93.81                \\
Reserved Mem (GB)  & 128    & 134                   & 136.78               \\
Peak Memory (GB)   & 139    & 136 ($\downarrow$3)   & 139                  \\
Frames             & 233    & 233                   & 257                  \\
Step Time (s)      & 62     & 56 ($\downarrow$10.7\%) & 68                   \\
Seq. Length        & 48k & 48k               & 52.8k ($\uparrow$10\%) \\
Batch Size         & 3      & 3                     & 3                    \\
Throughput         & 2,322  & 2,571 ($\uparrow$10.7\%) & 2,328                \\ \bottomrule
\end{tabular}
\end{table}

The quantitative results, summarized in Table \ref{tab:fuson_compare}, clearly illustrate significant gains in both throughput and load capacity. In our benchmark, the maximum sequence length supported by the GPU without operator fusion was capped at 48,000 tokens. Upon introducing operator fusion, the single-step execution time for the same load was significantly reduced from 62 seconds to 56 seconds, representing a 10.7\% throughput increase. A critical breakthrough is observed in the expansion of load boundaries: the optimized system can stably support sequence lengths up to 52,800 tokens---a 10\% improvement. Physically, this optimization saves approximately 3 GB of peak memory under identical loads.

To further dissect these system-level gains, we conducted a micro-level kernel benchmark as shown in Table \ref{tab:kernel_micro}. By analyzing the execution of a single \texttt{Fused AdaLN} operator across varying sequence lengths ($N$), we observe a consistent \textit{Pareto Improvement} in both computational and storage dimensions.

\begin{table}[htbp]
\centering
\caption{Performance and Memory Comparisonxx}
\label{tab:kernel_micro}
\small 
\setlength{\tabcolsep}{3pt} 
\begin{tabular}{lcccccccc}
\toprule
\textbf{N} & \multicolumn{3}{c}{\textbf{Forward (s)}} & \multicolumn{3}{c}{\textbf{Backward (s)}} & \multicolumn{2}{c}{\textbf{Mem (MB)}} \\
\cmidrule(lr){2-4} \cmidrule(lr){5-7} \cmidrule(lr){8-9}
(Seq) & \textbf{Fsd} & \textbf{Nat} & \textbf{Spd} & \textbf{Fsd} & \textbf{Nat} & \textbf{Spd} & \textbf{Fsd} & \textbf{Nat} \\ \midrule
8k  & 0.227 & 0.709 & 3.12$\times$ & 1.051 & 0.774 & 0.74$\times$ & 376  & 1072 \\
16k & 0.409 & 1.360 & 3.33$\times$ & 1.373 & 1.487 & 1.08$\times$ & 752  & 2012 \\
24k & 0.594 & 2.000 & 3.37$\times$ & 1.706 & 2.169 & 1.27$\times$ & 1125 & 2956 \\
32k & 0.783 & 2.645 & 3.38$\times$ & 2.067 & 2.878 & 1.39$\times$ & 1500 & 3942 \\
40k & 0.973 & 3.291 & 3.38$\times$ & 2.381 & 3.592 & 1.51$\times$ & 1877 & 4930 \\
48k & 1.161 & 3.933 & 3.39$\times$ & 3.320 & 4.256 & 1.28$\times$ & 2253 & 5912 \\
56k & 1.356 & 4.583 & 3.38$\times$ & 3.647 & 4.973 & 1.36$\times$ & 2626 & 6892 \\
64k & 1.547 & 5.240 & 3.39$\times$ & 3.986 & 5.668 & 1.42$\times$ & 3001 & 7878 \\ \bottomrule
\end{tabular}
\end{table}

\paragraph{Micro-architectural Analysis} In the forward pass, the fusion strategy achieves a stable speedup of \textbf{3.21$\times$ to 3.39$\times$}, effectively eliminating $\sim$70\% of computation time by transforming memory-bound operations into compute-bound ones. More importantly, in the backward pass, our proposed \textit{D-tile coalesced reduction} strategy demonstrates its scalability; as $N$ increases to 64,000, the backward speedup steadily grows to \textbf{1.42$\times$}, resolving the strided memory access bottleneck inherent in discrete gradient accumulation. 

From a storage perspective, the fusion kernel reduces activation memory by approximately \textbf{61.9\%} across all tested sequence lengths (e.g., at $N=64,000$, saving 4.8 GB). By ``collapsing'' the computational graph, intermediate tensors are intercepted within registers rather than written to HBM, providing a theoretical ``negative entropy'' for memory management. This micro-level efficiency directly translates to the 10\% expansion of global load limits observed in Table \ref{tab:fuson_compare}, allowing the model to bypass physical hardware constraints and process longer temporal sequences without relying on additional parallelization strategies.

\subsection{Training Stability and Convergence Verification}
While achieving substantial throughput gains, maintaining numerical stability and convergence quality is a fundamental prerequisite for any system-level optimization. We conducted rigorous monitoring of the training loss trajectories to ensure that our optimizations did not compromise the model's learning behavior.

\begin{figure}[t]
\centering
\includegraphics[width=\linewidth]{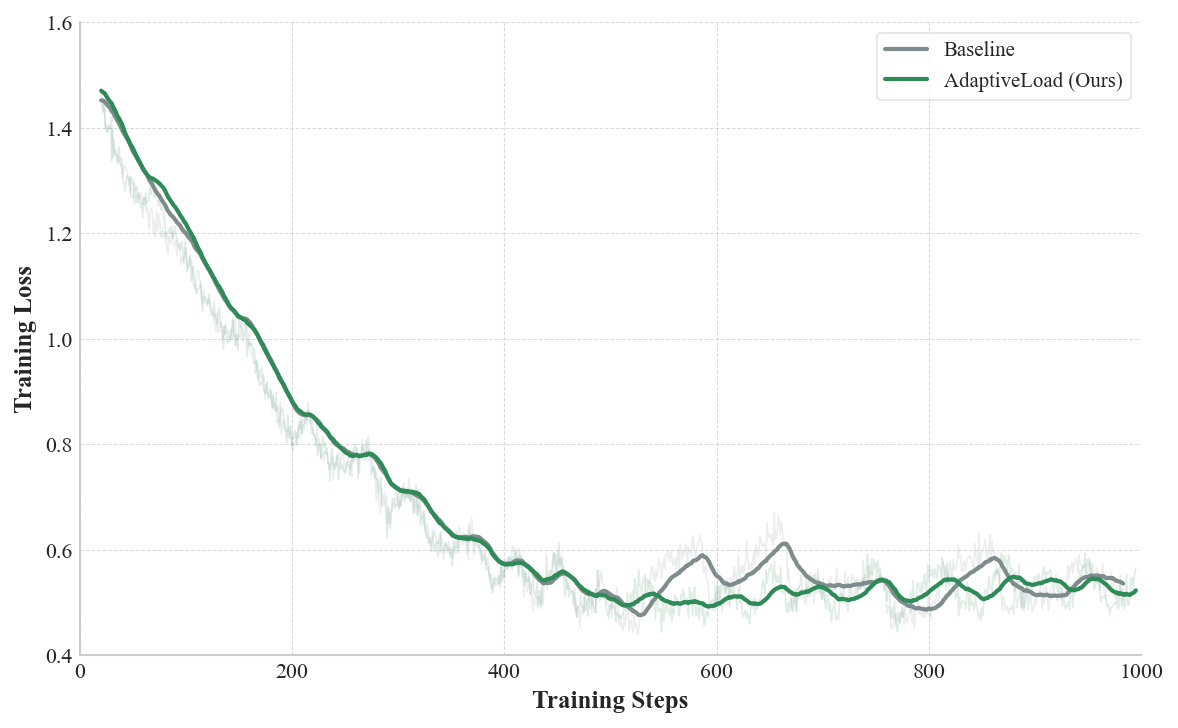}
\caption{Training loss curves of Wan 2.1 model for Baseline and AdaptiveLoad. The shaded curves represent raw loss values, while the solid lines show the smoothed trend.}
\label{fig:loss}
\end{figure}

As illustrated in Figure~\ref{fig:loss}, the loss curve of the AdaptiveLoad-optimized training remains highly congruent with the Baseline throughout the first 1,000 steps. This observation yields several critical conclusions:

\begin{itemize}
\item \textbf{Numerical Fidelity:} The highly overlapping trajectories validate that our D-tile coalesced reduction strategy, which utilizes float32 accumulation for critical gradient paths, preserves sufficient numerical precision to avoid divergence or accuracy degradation.
\item \textbf{Distribution Consistency:} The adaptive bucketing and dynamic batch size adjustment strategy does not disrupt the statistical distribution of the training data. Despite the intra-step re-alignment of sequences, the effective gradients across the cluster remain unbiased compared to the Baseline's stochastic sampling.
\item \textbf{Enhanced Robustness:} Interestingly, during the late training stage (approx. 500--900 steps), the AdaptiveLoad curve (dark green) exhibits slightly smoother convergence with fewer abrupt spikes compared to the Baseline. This suggests that by mitigating extreme sequence length imbalances, AdaptiveLoad may reduce the occurrence of "gradient noise" typically caused by highly skewed batch compositions.
\end{itemize}

The AdaptiveLoad framework seamlessly accelerates the training of the Wan 2.1 model while perfectly preserving its convergence robustness and final model quality.



\section{Conclusion and Future Work}
This paper presents \textit{AdaptiveLoad}, a full-stack optimization framework that addresses computational imbalance and memory access inefficiencies in large-scale video diffusion Transformer training. By implementing a dual-constraint adaptive load balancing mechanism driven by $O(S^2)$ complexity modeling, we reduce the load Coefficient of Variation (CV) by nearly 50\%, effectively eliminating synchronization bottlenecks. Complementing this, our fused LayerNorm-Modulate CUDA kernel—leveraging D-tile coalesced reduction—boosts training throughput by 28.63\% on a 16-GPU cluster while perfectly maintaining the numerical stability of the Wan 2.1 model. 

Future research will focus on scaling these optimizations to thousand-GPU clusters by integrating with hybrid parallelism strategies and generalizing cost-fitting models for emerging architectures like State Space Models (SSMs). Furthermore, we aim to utilize compiler technologies such as OpenAI Triton for automated kernel generation and incorporate real-time power monitoring to pursue a more power-efficient and sustainable approach to large-scale model training.

\bibliography{example_paper}
\bibliographystyle{icml2026}

\newpage
\appendix
\onecolumn



\end{document}